\documentclass[conference]{IEEEtran}
\IEEEoverridecommandlockouts
\usepackage{cite}
\usepackage[hidelinks]{hyperref}
\usepackage{amsmath,amssymb,amsfonts}
\usepackage{algorithmic}
\usepackage{graphicx}
\usepackage{textcomp}
\usepackage{xcolor}
\usepackage{mathtools}
\usepackage{booktabs}
\usepackage{siunitx}
\usepackage[export]{adjustbox}
\usepackage{balance}
\usepackage{subcaption}

\begin{document}

\title{Spatio-temporal Dynamics of Cellular V2X Communication in Dense Vehicular Networks}

\author{Behrad Toghi$^*$, Md Saifuddin$^*$, M. O. Mughal$^*$, Yaser P. Fallah$^*$\\
$^*$Connected and Autonomous Vehicle Research Lab (CAVREL)\\
University of Central Florida, Orlando, FL\\
\{toghi, md.saif, ozair\}@knights.ucf.edu, yaser.fallah@ucf.edu

}

\maketitle

\begin{abstract}
Cellular Vehicle-to-everything (C-V2X) communication is a major V2X solution proposed and developed by the 3rd Generation Partnership Project (3GPP). Our previous work has studied scalability aspects of C-V2X and demonstrated its potential for accommodating large numbers of vehicles in dense vehicular scenarios. However, existing studies in the scientific literature mostly have a network-level approach to the problem and do not assess the temporal and spatial dynamics of C-V2X networks in heavy network load situations. In this work we shed light on the spatio-temporal characteristics of these networks and investigate the effectiveness of the congestion control algorithm in dense vehicular ad-hoc networks (VANETs) in terms of settling time, stability, and reliability to be employed for the purpose of safety-critical vehicular applications, where latency plays a major role.
\end{abstract}
\section{Introduction}
In 1999, the Federal Communications Commission (FCC) of the United States specified 75 MHz of spectrum in the 5.9 GHz band for vehicle safety applications and intelligent transportation system (ITS). Dedicated Short Range Communications (DSRC) technology based on IEEE 802.11p, IEEE 1609.X, and SAE J2945 standards \cite{yfallah:idmtvt} and Cellular Vehicle-to-everything (C-V2X) \cite{bazzi2019survey, eckermann2019performance}. Albeit the fact that DSRC is considered as the current primary solution for the vehicular communications, many contributors, e.g., regulators, automakers, and tier-1 suppliers have shown interest in C-V2X networks as an alternative. These vehicular ad-hoc networks (VANETs) are expected to enable vehicles to share their position and mobility information in the form of either basic safety messages (BSMs) or model-based communication \cite{yfallah:mbcsyscon}, as recently proposed in \cite{hnmahjoub:cavs, hnmahjoub:vtc, hnmahjoub:syscon}.

In order to address the reliability and latency requirements of vehicle-to-everything (V2X) communication, the 3GPP release 14 standard specified Mode-3 and Mode-4 operation modes. While Mode-3 enables the in-coverage communication among user equipments (UEs), Mode-4 specifically focuses on the ad-hoc out-of-coverage communication where the resource allocation has to be realized in an unsupervised and distributed fashion. Mode-4 utilizes an enhanced packet collision avoidance mechanism based on channel occupancy to cut down the collision probability and improve communication latency and reliability. Our focus in this study is on Mode-4 sidelink communication \cite{3gpp:36213, 3gpp:36885}, to which we refer as C-V2X for the sake of readability.

Among the recent studies on the performance of C-V2X communication, authors in \cite{jgozalvez:vtc2017spring} explored different types of transmission errors, such as errors due to propagation loss, packet collisions, and half-duplex operation, using system level simulations. Authors in \cite{bazzi2019survey} present a detailed survey on perspectives and contrasts between C-V2X and DSRC. Recently an open-source C-V2X network simulator is released by \cite{eckermann2019performance}. Authors in \cite{zhao2019cluster} propose a novel method for resource allocation in C-V2X networks based on clustering. One of the few studies on congestion control for C-V2X communication is presented by authors in \cite{mansouri2019first}. In \cite{bkang:cv2xPowerControl}, a power control scheme is proposed which improves the performance via reducing the transmission power in dense networks.

In this article, which supplements our previous work \cite{btoghi:vtc2019, btoghi:vnc}, we investigate the spatial and temporal characteristics of the C-V2X communication with and without enabling the distributed congestion control (DCC) mechanism, as suggested in \cite{sae:j2945}. This study helps us to further understand the effectiveness and capabilities of C-V2X in the context of safety-critical vehicular applications. In particular, considering a vehicular ad-hoc network (VANET) as a dynamic system, we are interested in settling-time and temporal behavior of this system while adapting to a new environment, as well as the spatial characteristics of the interaction among UEs in the network. We believe, such approach to the problem is absent in the previous related works, as they mostly rely on the network-level metrics, such as packet error rate, which does not necessarily reflect any insight about spatio-temporal statistics of the network. 
\section{DCC Enabled C-V2X Communication}
In what follows, we provide a brief overview of the PHY and MAC of C-V2X communication, followed by details of the distributed congestion control algorithm proposed in \cite{sae:j2945} and investigated in \cite{btoghi:vtc2019, mansouri2019first}. C-V2X utilizes similar time and frequency divisions to those of the LTE technology \cite{3gpp:36213}, i.e., \emph{subframes} in time domain and \textit{subchannels} in frequency domain. A group of RBs form a candidate single-subframe resource (CSR), hereafter referred to as a \emph{radio resource}. Sensing-based Semi-persistent Scheduling (SB-SPS) \cite{3gpp:36213} is utilized in C-V2X Mode-4 communication as the distributed scheduling mechanism. Every UE senses the channel and keeps track of the received signals from its neighboring nodes for the last $1000 ms$ (sensing window). For a packet generated at time $n$, the UE utilizes information from the last sensed window to schedule a set of transmission opportunities in the future.
 
Congestion control algorithms are aimed at avoiding heavy performance degradation in dense network, in our case dense VANETs. This goal can be realized by either reducing the rate of message broadcast or by limiting the communication range of UEs \cite{yfallah:idmtvt, sae:j2945, btoghi:vtc2019}. UEs broadcast BSMs at a rate of 10 Hz in the baseline operation mode \cite{sae:j2735}. Transmitting messages at this rate in high-density scenarios overloads the network and negatively impacts the communication reliability. Rate control algorithms are designed \cite{gbansal:limerictvt, yfallah:tits2011} to reduce the load created by every node on the network while maintaining the required latency. Rate control algorithms govern the inter-transmit time (ITT) between the packets based on the vehicle density in each vehicle's proximity. Alongside with probing the vehicle density and controlling ITT, each vehicle evaluates the Position Tracking Error (PTE). PTE measures the precision of the location estimation of the surrounding vehicles' estimated by the host vehicle. The host vehicle broadcasts a message regardless of the calculated ITT if $\text{PTE}>50cm$. 

UEs to control their communication range and avoid excessive packet collisions based on channel busy percentage (CBP), a metric for probing the network utilization. SUPRA algorithm, proposed in \cite{yfallah:tvtsupra2016}, maps CBP to the radiated power of each UE in order to control CBP in the range enforced by the system requirements in the application layer.
\begin{figure}[t]
\includegraphics[width=.48\textwidth, left]{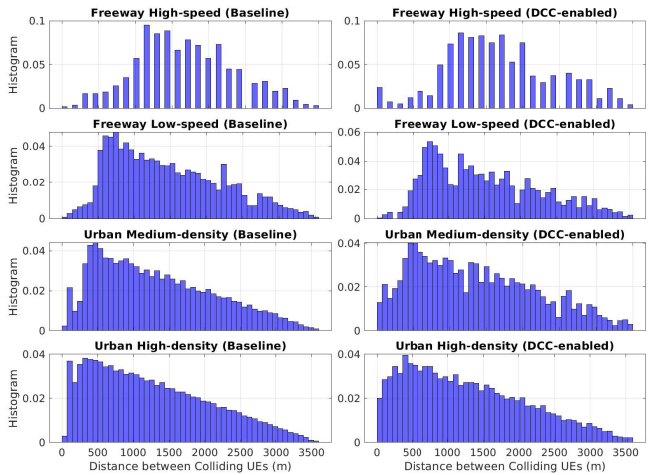}
\caption{Probability of packet collision between UEs versus distance in different node densities. \textit{(Left)} Baseline C-V2X, \textit{(Right)} DCC-enabled C-V2X.}
\label{fig:fig1}
\end{figure}
\section{Simulation Setup}

We employ a realistic propagation channel model suggested in \cite{ghayoor:emulator, btoghi:vnc}. The Fowlerville channel model is derived from a large data set, collected during field trials on FTT-A Fowlerville Proving Ground by the Crash Avoidance Metrics Partnership (CAMP) consortium in collaboration with USDoT. We study a network of moving vehicles in a $3.6 km$ straight highway with total 12-lanes in two directions. Other simulation assumptions and configurations are listed in Table \ref{table:configs}.
\begin{table}[b]
\centering
\caption{Simulation Parameters \& Configurations}
\begin{center}
\bgroup
\def\arraystretch{1.4}
\begin{tabular*}{0.37\textwidth}{@{\extracolsep{\fill} }  l r l r  }
\hline
\hline
Time $(T_{sim})$     & 120 s          & $P_{\text{min}}$                & 10 dBm\\
Payload Size         & 190 B          & $P_{\text{max}}$                & 23 dBm\\
MCS Index            & 5              &$\text{ITT}^{\text{max}}$        & 600 ms\\
Carrier Freq.        & 5860 MHz       & $CBP_{\text{min}}$              & 50\% \\
Bandwidth            & 10 MHz         & $CBP_{\text{max}}$              & 80\% \\
\hline
\end{tabular*}
\egroup
\label{table:configs}
\end{center}
\end{table}
\begin{figure}[t]
\includegraphics[width=.48\textwidth, left]{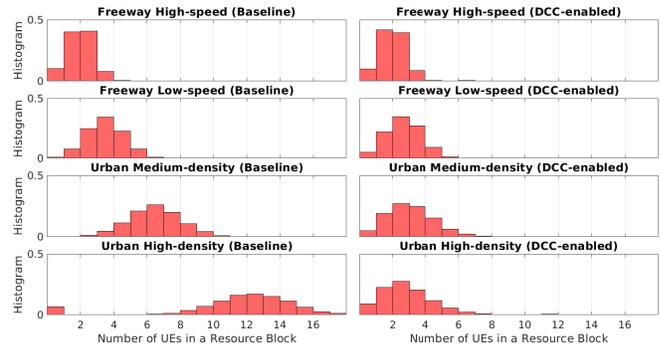}
\caption{Distribution of the number of UEs occupying a resource block (RB).}
\label{fig:fig3}
\end{figure}
We consider 5 scenarios with different node densities and vehicle velocities as suggested in \cite{3gpp:36885}. \textit{Freeway High-speed} scenario contains 300 vehicles with density of 7 Veh/(km.lane)\footnote{Vehicle per kilometer per lane}, cruising at 140km/h. \textit{Freeway Low-speed} scenario contains 600 vehicles with density of 14 Veh/(km.lane), cruising at 70km/h. \textit{Urban Medium-density}, \textit{High-density}, and \textit{Ultrahigh-density} scenarios accommodate 1200 (28 Veh/(km.lane)), 2400 (56 Veh/(km.lane)), and 4800 (111 Veh/(km.lane)) vehicles respectively, all moving at 15km/h.

\begin{figure}[t]
\includegraphics[width=.48\textwidth, left]{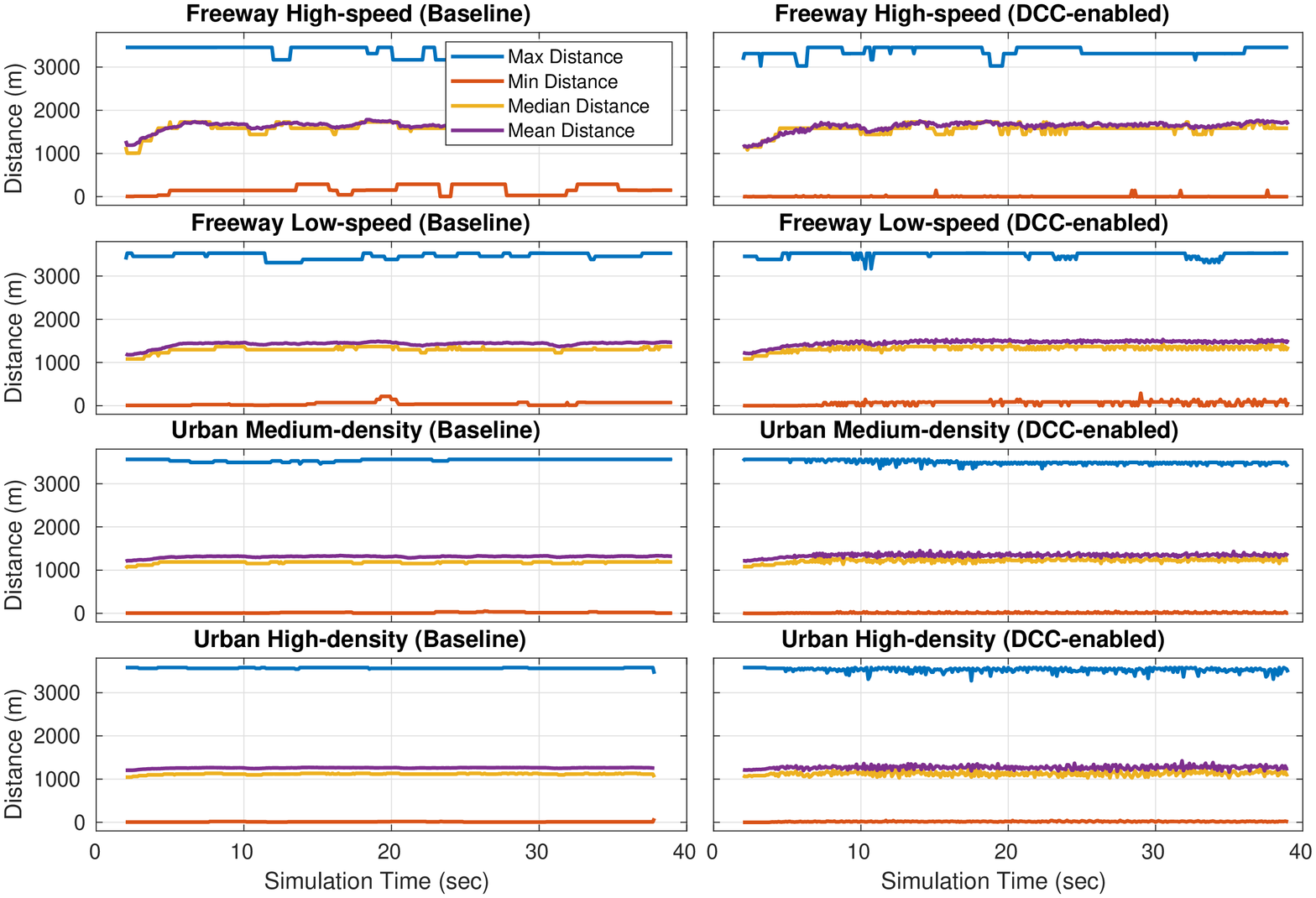}
\caption{Temporal behavior of statistics of the geographical distance between colliding UEs.}
\label{fig:fig4}
\end{figure}

\section{Analysis and Results}

In this work we are aiming to provide a new point-of-view, compared to the existing studies which study the performance of C-V2X communication. These works mainly evaluate the performance \cite{btoghi:vnc,jgozalves:msn, jgozalvez:vtc2017spring} and scalability \cite{btoghi:vtc2019} of V2X systems in terms of network-level key performance indicators (KPIs), e.g., packet error rate, inter-packet gap, network throughput, and information age. However, in the context of vehicular safety applications we are specifically interested to investigate how \textit{fast} the nodes are able to adapt to a dynamic environment and how \textit{reliable} is the communication with vehicles in their close proximity. While the former is important as communication latency can lead to high-risk driving situations or even crash scenarios, we are interested in the latter since reliability in close proximity communication is much more crucial than that among nodes placed far from each other.

As stated in previous works \cite{jgozalvez:tvt2018, btoghi:vnc}, high-density VANETs experience high loads of safety messages and consequently excessive packet drops. Looking solely at a network-level metric, such as packet error rate, does not provide an insight on the spatial and temporal dynamics of the network, e.g. settling-time and geographical distance between conflicting UEs. Hence we introduce a new point-of-view for the purpose of evaluating C-V2X networks in highly-congested vehicular scenarios. Additionally, we also compare the baseline operation of C-V2X \cite{3gpp:36213} to the DCC-enabled C-V2X \cite{btoghi:vtc2019} to further study the impact of DCC on the spatio-temporal characteristics of the network. We conclude this section with an important observation on inter-dependency of the MAC layer mechanisms and the rate control of DCC-enabled C-V2X.

The design rationale behind the multiple access mechanism employed in C-V2X Communication, i.e., SB-SPS, is to sense and keep a history of the communication channel and utilize it to avoid conflicts, e.g., packet collision among nodes in close proximity. This purpose is realized through exempting radio resources with high received signal strength, as described in \cite{btoghi:vnc, 3gpp:36331}, from the resource pool of every UE. In the case of baseline operation, where all UEs transmit their packets with a constant radiated power, the mapping between the received signal strength (RSS) and distance of the transmitter depends on the characteristics of the radio propagation medium, mostly referred to as the propagation channel model. Estimating this mapping is a very challenging task due to the stochastic nature and unpredictable environment of the problem. Hence, one may question the effectiveness of the SB-SPS algorithm in mitigating packet collisions among neighboring UEs, and on the other hand, the impact of range control in DCC-enabled C-V2X networks on the above-mentioned matter is also uncertain.

Thus, we measure the geographical distance between conflicting UEs, i.e., UEs that have selected the exact same radio resource for their packet transmission. Fig. \ref{fig:fig1} illustrates the distribution of these measured values normalized by the number of UEs that have placed in that distance. In other words, this figure shows the probability, $\mathbb{P}(d_{Rx}^{Tx})$, of experiencing a conflict between two nodes with distance $d_{Rx}^{Tx}$. Results in Fig. \ref{fig:fig1} state that SB-SPS effectively brings down the probability of packet collision among vehicles in close proximity. From a vehicle safety point-of-view, this observation contains crucially important insights. As a matter of fact, most of the near-crash and safety-critical scenarios involve vehicles with relative distance of less than 100m hence it is essential to ensure the communication reliability, particularly among vehicles that are placed in the immediate or close neighborhood of each other.
\begin{figure}[t]
\includegraphics[width=.48\textwidth, left]{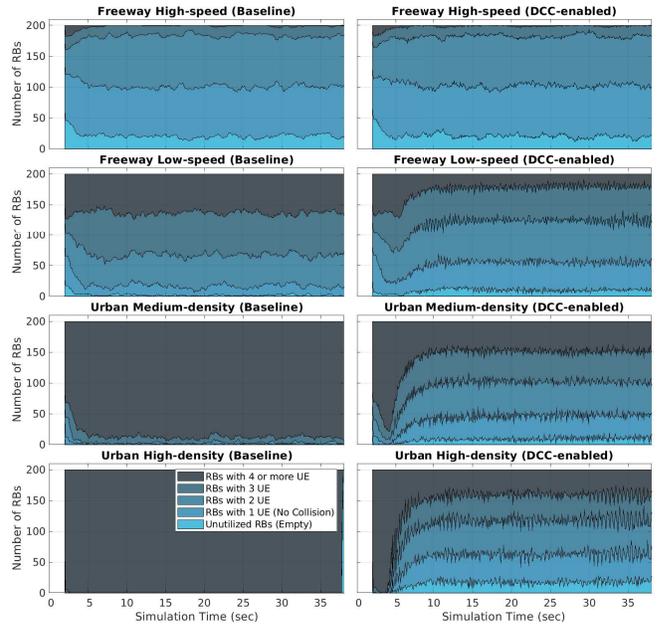}
\caption{Number of UEs per resource block (RB) versus time.}
\label{fig:fig5}
\end{figure}

As discussed in \cite{btoghi:vtc2019}, DCC maintains the communication range via governing the radiated power, hence, we are interested to investigate the inter-operability of power control and SB-SPS. Fig. \ref{fig:fig1} demonstrates a comparison between the baseline and DCC-enabled C-V2X. It can be inferred that in very near distances ($<$100m), SB-SPS algorithm in DCC-enabled mode is not as effective as it is in the baseline mode. This is due to the fact that the MAC layer sensing threshold ($Th_{\text{SPS}}$) is set independent of the radiated power, thus, SB-SPS falsely estimates the distance between two close nodes based on the measured RSS value. This issue can be addressed by varying $Th_{\text{SPS}}$ in an adaptive fashion which can partially avoid the observed problem. The majority of packet collisions happen among nodes that are placed in far distances and only less than 5\% percent of the conflicts are in vehicles within the range of 500m.

The resource allocation mechanism is designed to avoid excessive and continued packet collisions, if possible. This can be translated to maximizing the channel utilization and minimizing the packet error rate. Based on our simulation setup (discussed in Section III), every selection window contains 200 radio resources to be chosen by each UE. As an instance, the Freeway high-speed scenario consists of a fully connected network of 300 nodes. Thus, the Pigeonhole principle enforces that, ideally, no RB should be chosen by more than one UE. However, due to the imperfections in the distributed resource allocation, such results is not observed in Fig. \ref{fig:fig3}, in which some RBs are chosen by more than 4 UEs while some other RBs being unutilized. This issue becomes even more noticeable in the denser Urban High-density scenarios, where DCC-enabled C-V2X performs significantly better, as also shown in \cite{btoghi:vtc2019}.

Fig. \ref{fig:fig4} and \ref{fig:fig5} illustrate the time behavior of the above-mentioned perspectives. Such temporal investigation is important since the V2X systems mostly involve time-critical safety applications. Hence, it is crucial for a safety V2X system to adapt to a new environment in a timely manner. Stability, on the other hand, is another important requirement since inconsistent and undesired behavior of vehicular safety systems can be catastrophic. Fig. \ref{fig:fig4} shows the similar stable temporal behavior for both baseline and DCC-enabled modes, in which both settle around 5s after the simulation initiates. This is a relatively long duration of time in the scale of near-crash scenarios and, we believe, can be improved by changing the configurations of SB-SPS. It is also worth mentioning that our tests in this study have been conducted in an extreme case, in which vehicles suddenly meet at a point without any prior knowledge or situational awareness; such case is not likely to happen in a real-world scenario. Fig. \ref{fig:fig5} conveys two messages, settling time for the SB-SPS and DCC algorithms. Comparison between DCC-enabled and baseline C-V2X shows the significant effect of DCC on the distribution of the number of conflicting nodes per RB.

\begin{figure}[t]
\includegraphics[width=.48\textwidth, left]{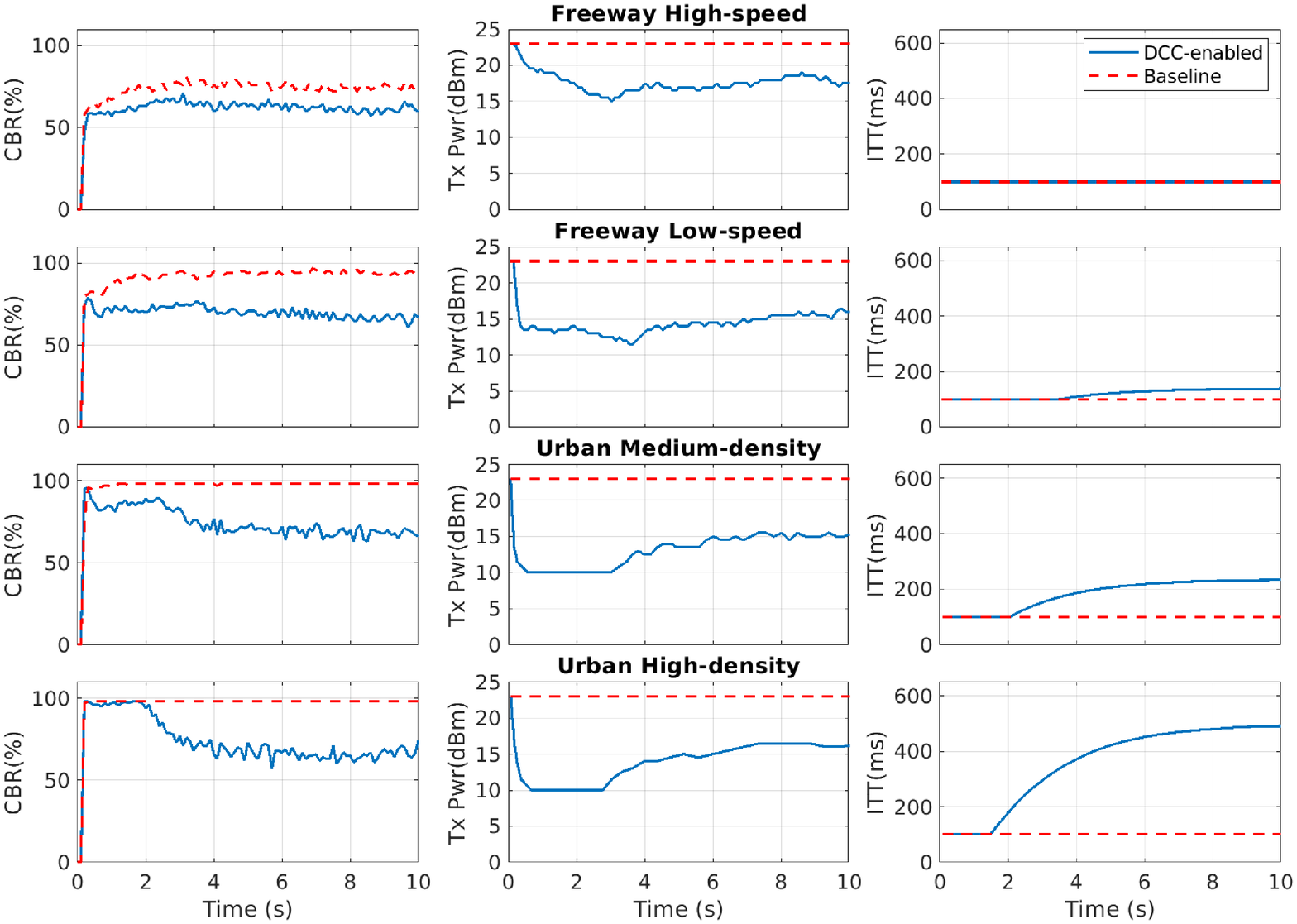}
\caption{Congestion Control parameters versus time. Transmission Rate and radiated power are shown as well as their impact on channel utilization.}
\label{fig:fig6}
\end{figure}

We also investigate the temporal behavior of the DCC parameters, i.e., Message Rate and Radiated Power, as well as their impact on the Channel Busy Percentage (CBP). Fig. \ref{fig:fig6} contains a very important observation which shows the impact of Range and Rate control in DCC algorithm. As it can be inferred from the algorithms in \cite{btoghi:vtc2019} that the range control algorithm has a much quicker response to change in environment conditions than that for the Rate Control. This is consistent with our observation in Fig. \ref{fig:fig6} where comparing CBP, Transmission Rate, and Radiated Power shows us the important impact of Rate Control while the Range control does not add a significant improvement.

\begin{figure}
    \centering
    \begin{subfigure}[b]{0.22\textwidth}
        \includegraphics[width=\textwidth]{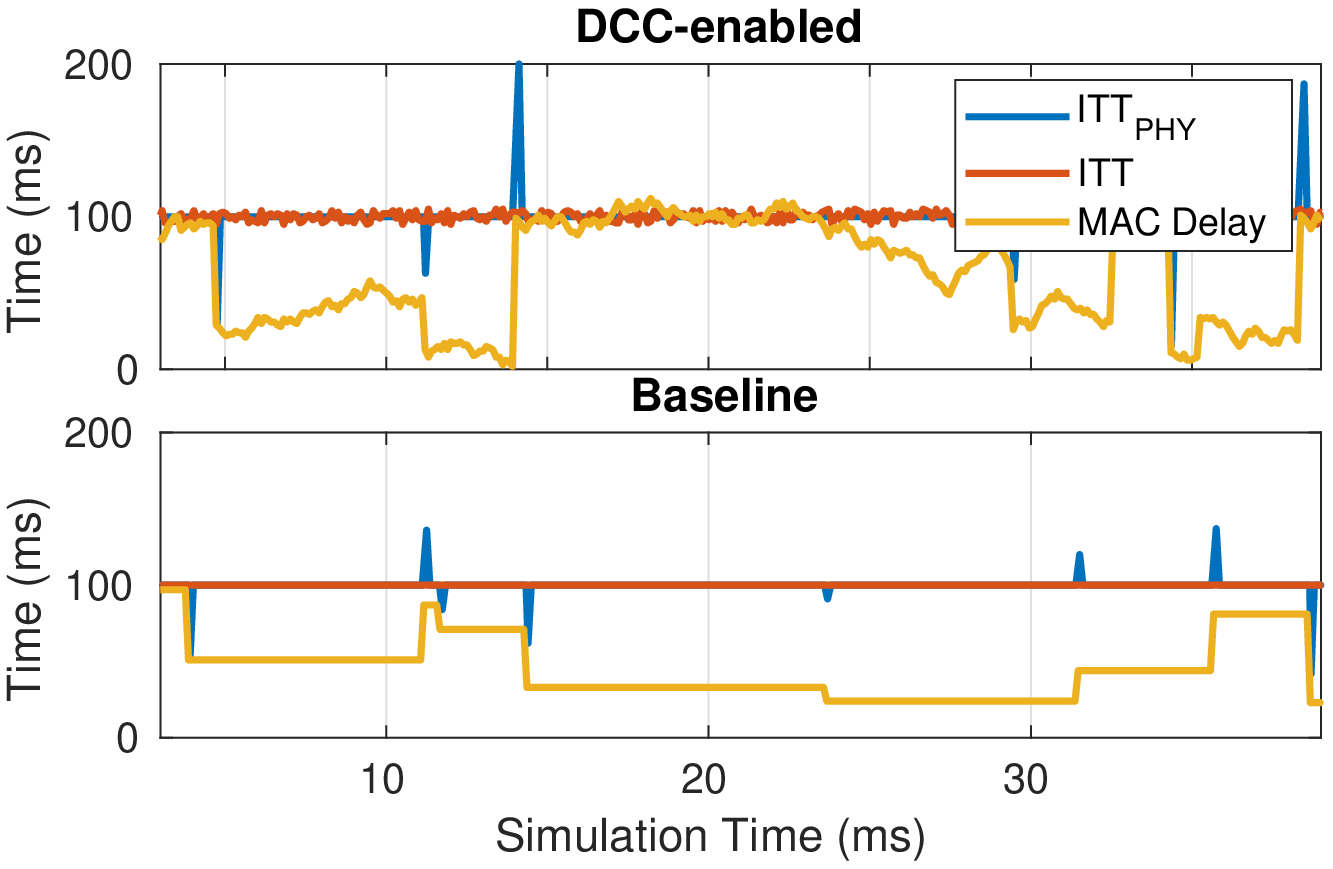}
        \caption{Freeway High-speed}
        \label{fig:fig7a}
    \end{subfigure}
    ~ 
    \begin{subfigure}[b]{0.22\textwidth}
        \includegraphics[width=\textwidth]{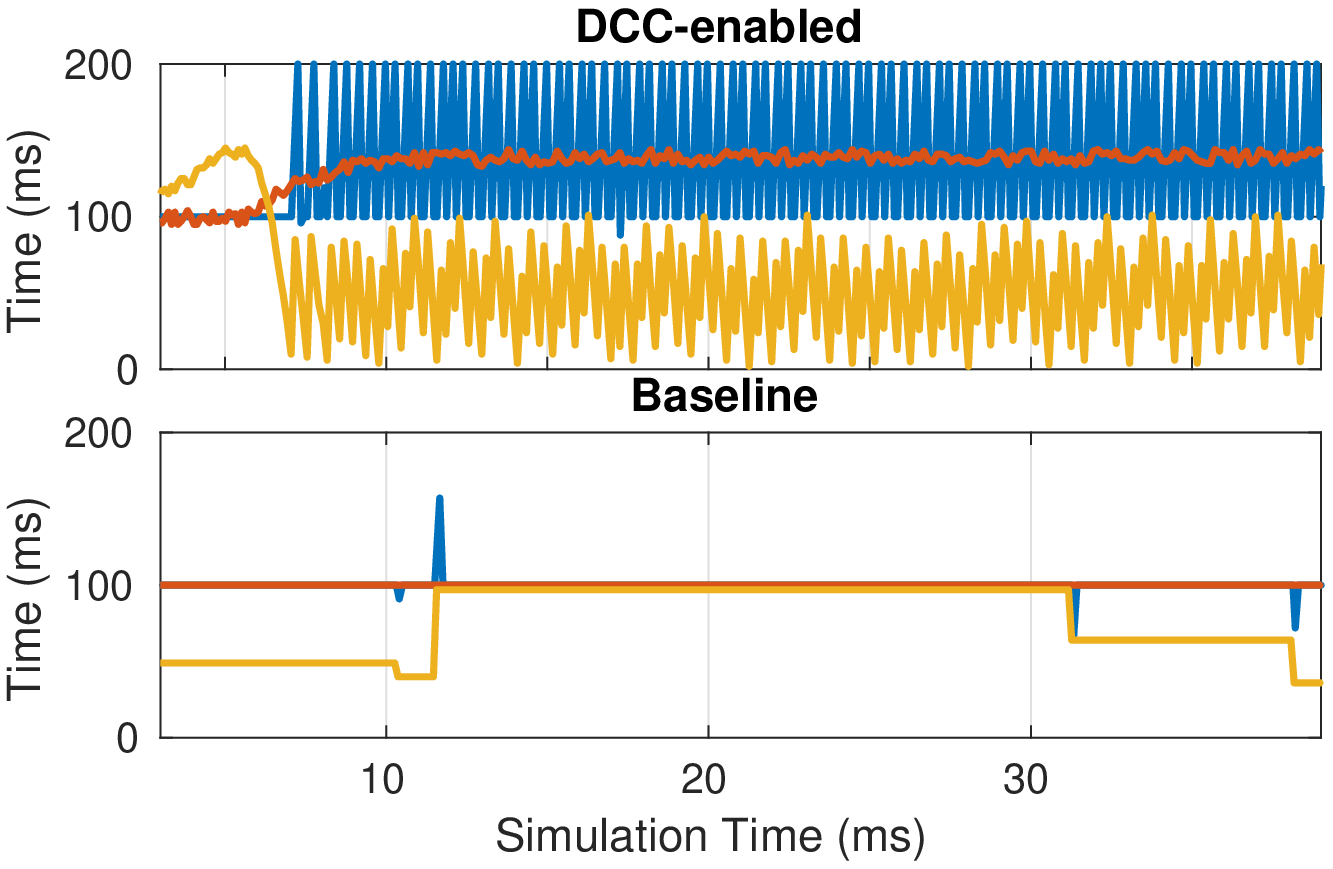}
        \caption{Freeway Low-speed}
        \label{fig:fig7b}
    \end{subfigure}
    
    \begin{subfigure}[b]{0.22\textwidth}
        \includegraphics[width=\textwidth]{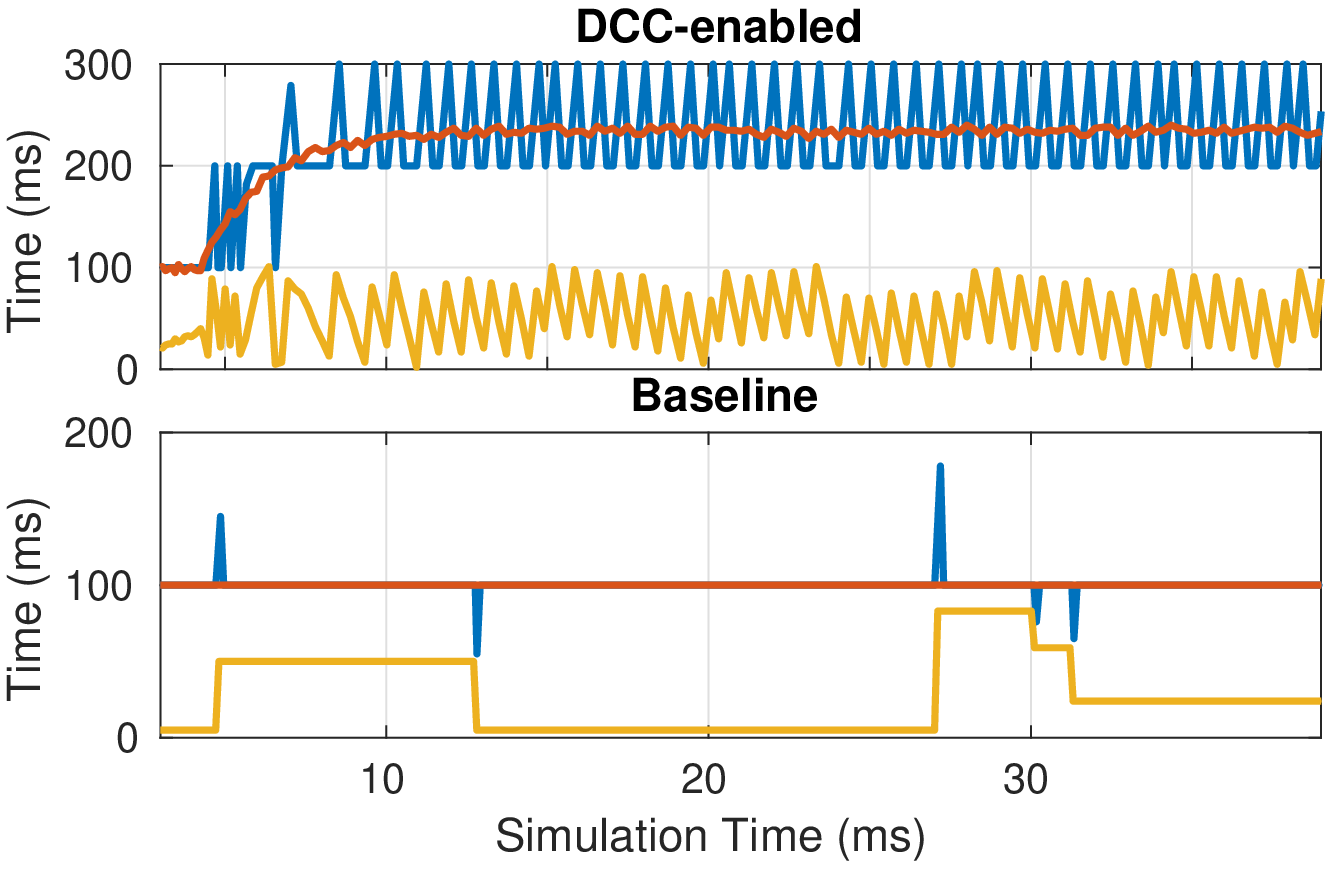}
        \caption{Urban Medium-density}
        \label{fig:fig7c}
    \end{subfigure}
    ~
    \begin{subfigure}[b]{0.22\textwidth}
        \includegraphics[width=\textwidth]{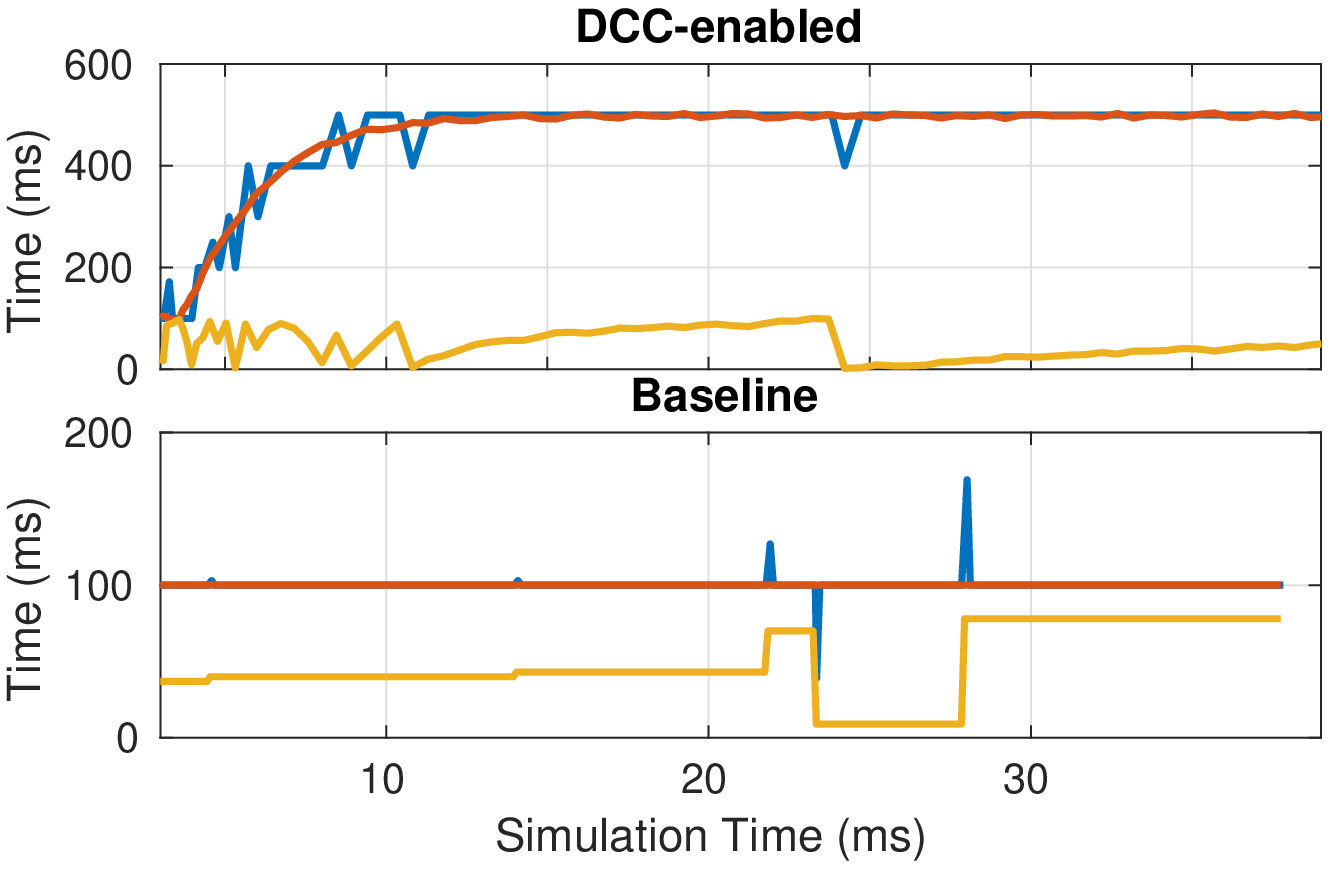}
        \caption{Urban High-density}
        \label{fig:fig7d}
    \end{subfigure}
    
    \caption{Analysis of MAC delay and co-existence of SB-SPS with DCC Rate Control}
    \label{fig:macdelay}
\end{figure}

Additionally, we inspected the delay between packet generation time in application layer and transmission time in PHY layer, i.e., MAC delay (Fig. \ref{fig:macdelay} \textit{yellow}). Results show that MAC delay varies between $0$ and $100ms$. In order to understand the reason behind this fluctuations, we illustrate the time gap between two consecutive packets transmitted in the PHY layer, i.e., $\text{ITT}_{\text{PHY}}$, and compare it to ITT (Fig. \ref{fig:macdelay} \textit{blue} and \textit{red}). MAC layer receives a packet from higher-layers and schedules it in one of the reserved radio resources. These transmission opportunities are separated by $100ms$ time intervals, thus, if the packet misses a transmission opportunity, it needs to wait for the next on-coming reservation. In other words, $\text{ITT}_{\text{PHY}}$ fluctuates between the closest multiples of $100ms$ to ITT, as shown in Fig. \ref{fig:macdelay}. The conclusions from this discussion are, Rate Control and SB-SPS mechanisms are able to co-exist and operate jointly and SB-SPS mechanism is capable  of handling variable transmission rate in DCC-enabled C-V2X.
\section{Concluding Remarks}
The Cellular Vehicle-to-everything (C-V2X) technology has gained strong interest as an alternative for the existing IEEE802.11p or Dedicated Short Range Communication (DSRC). Both of these V2X solutions need sophisticated congestion control algorithms in order to be able to operate in high density vehicular scenarios. Previous works have analyzed the efficiency and performance of C-V2X communication, equipped with the Distributed Congestion Control (DCC) algorithm in terms of network-level metrics, such as packet error rate and inter-packet gap. In this work, we present a new perspective to the problem and investigate the temporal and spatial dynamics of the network, to study settling time, adaptability, stability, and effectiveness of V2X communication for safety-critical applications. Our results demonstrate the effectiveness of resource allocation of the C-V2X communication, in addition to the significant enhancement resulted by the congestion control algorithm in terms of communication latency and reliability.
\balance

\bibliography{refs.bib}{}
\bibliographystyle{unsrt}
\end{document}